# Analysis of the Formation of Structures and Chaotic Dynamics in a Nonrelativistic Electron Beam with a Virtual Cathode in the Presence of a Decelerating Field

E. N. Egorov, Yu. A. Kalinin, A. A. Koronovskii,
Yu. I. Levin, and A. E. Hramov

**Abstract**—A nonrelativistic electron beam with a virtual cathode situated in the diode gap with a decelerating field is experimentally and theoretically studied. A 1D model of the electron beam in the presence of a decelerating field is constructed. Nonlinear nonstationary processes in this system are investigated by means of numerical analysis of the model. The processes are described and interpreted with regard to formation and interaction of structures in the electron flow. The theoretical results are qualitatively confirmed by the experimental data showing that the system under study can be considered as a promising controlled source of chaotic oscillations in the microwave range.

PACS numbers: 05.45.Pq

## INTRODUCTION

The analysis of the complicated nonlinear dynamics and physical processes in high-intensity electron beams with a virtual cathode (VC) is an extremely important and topical problem. The significance of this problem has been indicated in a substantial number of related works (see, for example, [1–10]). This study is important for the fundamental investigation processes governing formation of the chaotic dynamics and structures in the beam–plasma systems [5, 6, 8, 9, 11–13], for the applied analysis related to the development and optimization of the VC-based high-power microwave oscillators (vircators) [4, 5, 8, 14], for acceleration of charged particles with an oscillating VC [15, 16], and for creation of the sources of broadband noiselike radiation with different power levels [10, 17].

Note that the experimental study and realization of the VC-based oscillators is a very complicated problem involving application of high-intensity relativistic electron beams with currents higher than the limiting vacuum current (supercritical current) [18]. In this case, one cannot perform a detailed study of the characteristics of generation of vircator systems and physical processes in the electron beam with a VC. One of the possible methods for easing the experimental conditions for generation of microwave oscillations with the help of a VC involves application of systems with an additional deceleration of electrons in which a nonstationary oscillating VC is formed owing to a strong deceleration of the beam [10, 17, 21].[1] Such a system can be characterized by formation of a VC and generation of a chaotic broadband signal at low currents and densities of the electron beam. Hence, it is of interest to perform an in-depth experimental study of physical processes in a nonrelativistic beam with a VC via conventional approaches of the physical experimentation of vacuum nonrelativistic microwave electronics [22].

In this case, it is expedient to study nonstationary nonlinear processes in a nonrelativistic electron beam with a VC formed in the presence of a decelerating field. A plane diode gap transmitting an electron flux in the case when a decelerating voltage is applied to the second (output) grid can be considered as the base model. Such a model of the plane diode gap is one of the base models that are widely used for the theoretical and numerical analysis of processes in the flows of charged particles with a VC formed in the equipotential space (in the absence of an additional deceleration) [9, 13, 23–25].

The purpose of this study is to numerically investigate nonstationary nonlinear processes, including the dynamic chaos and the formation and interaction of coherent structures in a nonrelativistic electron beam with a VC in the plane diode gap with a decelerating field, on the basis of the method of large particles. The theoretical results on the chaotic VC oscillations in the presence of a decelerating field are verified on a specially designed prototype in which a nonstationary VC is formed in a nonrelativistic electron beam in the diode gap with a decelerating potential. Note that, in the case of deceleration of an electron flux, such systems with a VC can serve as the sources of a noiselike broadband medium-power chaotic microwave signal [10]. Therefore, the theoretical and experimental analysis of such systems is of great practical importance.

---

[1] Note a certain relation between oscillations in such an electron system with deceleration and oscillations in the Barkhausen–Kurz oscillator [19] and oscillations in an electron–wave oscillator with a decelerating field ([20, Chapter 5]).

## 1. MODEL AND SCHEME FOR THE NUMERICAL SIMULATION

To study the electron flux with a VC formed in the presence of a decelerating field, we consider a model of a plane diode gap in which a nonrelativistic electron beam is injected through the first (input) grid. The decelerating field is generated owing to the presence of a decelerating voltage at the second (output) grid of the diode gap.

In the absence of deceleration, the electron flow is completely transmitted by the diode gap (stationary transmission of the beam in the absence of a VC). However, a nonstationary oscillating VC may emerge in such a system in the presence of a decelerating field in a certain range of the potential of the second grid. The VC oscillations can be used for generation of microwave radiation. In contrast to conventional vircators, which employ the relativistic beams and whose pulse power ranges from several tens to several hundreds of megawatts [13], these oscillations are characterized by a significantly lower power. (In [17], the integral continuous-wave power of the oscillations in a nonrelativistic beam with a VC formed in the presence of a decelerating field is a few watts for the centimeter wave range.)

Consider the scheme for the numerical simulation of nonlinear nonstationary processes in a nonrelativistic electron flux that is used in this study. We assume the 1D motion of the electron beam. In the case of planar configuration of the diode gap, the electron flux represents a set of large particles (charged sheets) injected at equal time intervals with a constant velocity into the interaction space. Instead of dimensional potential $\varphi$, intensity $E$ of the space-charge electric field, electron density $\rho$, electron velocity $v$, and spatial $x$ and time $t$ coordinates, we will use the following dimensionless variables:

$$\varphi = (v_0^2/\eta)\varphi', \quad E = (v_0^2/L\eta)E', \quad \rho = \rho_0\rho',$$
$$v = v_0 v', \quad x = Lx', \quad t = (L/v_0)t'. \quad (1)$$

Here, primes (omitted below) denote dimensionless variables, $\eta$ is the specific electron charge, $v_0$ and $\rho_0$ are the static (unperturbed) velocity and density of the electron flux at the entrance to the system, and $L$ is the length of the drift space.

For each charged sheet (large particle), we solve the nonrelativistic equations of motion, which can be represented in dimensionless variables (1) as

$$d^2x_i/dt^2 = -E(x_i), \quad (2)$$

where $x_i$ is the coordinate of the $i$th charged sheet and $E(x_i)$ is the intensity of the space-charge electric field at the point with coordinate $x_i$.

For the calculation of the space-charge field intensity and potential and the charge density, we employ a uniform spatial mesh with step $\Delta x$. The potential of the space-charge field is determined from the Poisson equation, which is represented in the 1D approximation as

$$\partial^2\varphi/\partial x^2 = \alpha^2\rho(x). \quad (3)$$

Here, $\alpha = \omega_p L/v_0$ is the Pierce parameter [23], which represents the unperturbed electron drift angle with respect to plasma frequency $w_p$. In Eq. (3), we assume that the density of the positive ion background is relatively low and can be neglected in the simulation of oscillatory processes in the electron beam. Field intensity $E$ of the space charge is determined via numerical differentiation of potential $\varphi$ in the diode gap.

Poisson equation (3) must be supplemented with the following boundary conditions:

$$\varphi(x = 0) = \varphi_0, \quad \varphi(x = 1) = \varphi_0 - \Delta\varphi, \quad (4)$$

where $\varphi_0$ is the accelerating potential (in the above normalization, $\varphi_0 = 1$) and $\Delta\varphi$ is the decelerating potential difference between the grids.

To calculate the space-charge density, we employ the linear weighting of the particles (sheets) at the spatial mesh (method of particles in a cell), while decreasing the mesh noise [26]. In this method, the space-charge density at the $j$th mesh node (i.e., at the point with coordinate $x_j = j\Delta x$) is written as

$$\rho(x_j) = \frac{1}{n_0}\sum_{i=1}^{N}\Theta(x_i - x_j), \quad (5)$$

where $x_i$ is the coordinate of the $i$th particle, $N$ is the total number of large particles, $n_0$ is the calculation parameter equal to the number of particles per cell in the unperturbed state, and

$$\Theta(x) = \begin{cases} 1 - |x|/\Delta x, & |x| < \Delta x \\ 0, & |x| > \Delta x \end{cases} \quad (6)$$

is the piecewise-linear function that determines weighting of a large particle at the spatial mesh with step $\Delta x$.

Large particles that arrive at the output grid, reflect from the VC, and return to the first (input) grid are absorbed at the boundaries.[2]

The main parameters of the numerical scheme (such as number of nodes $N_m$ of the spatial mesh and number $n_0$ of particles per cell in the unperturbed state) are $N_m = 800$ and $n_0 = 24$. These values correspond to a number of particles in the calculation space in the unperturbed state of $N = 19\,200$. The parameters of the numerical scheme are chosen so as to reach the desired accuracy and ensure adequacy of the results for the analysis of complicated nonlinear processes, including the deterministic chaos in the electron–plasma system under study [26, 27]. The equations are solved with the use of the step-over scheme [26] with the second order of accuracy. The Poisson equations are integrated via the error-vector propagation method [28].

## 2. NONLINEAR VIRTUAL-CATHODE DYNAMICS DURING A VARIATION IN THE CONTROL PARAMETERS

Consider the results of numerical simulation of the nonlinear dynamics for an electron flux with a VC in the presence of a decelerating field.

Figure 1 demonstrates characteristic regimes of the electron-flux oscillations in the diode gap in the plane of such parameters as Pierce parameter $\alpha$ and the decelerating potential difference between the grids, $\Delta\varphi$. Figure 2 shows power spectra $P(f)$ and phase portraits (plotted by means of the delay method [29]) for the oscillations of the center of mass, $x_{cm}(t)$, of the electron beam in the diode gap for the Pierce parameter $\alpha = 0.9$ and various decelerating potential differences $\Delta\varphi$. The center-of-mass coordinate is determined as $x_{cm} = \frac{1}{N}\sum_{i=1}^{N} x_i$, where $x_i$ is the coordinate of the $i$th large particle and $N$ is the total number of particles in the interaction space. The center-of-mass oscillations are chosen as a parameter characterizing the behavior of the system in accordance with [11]. Note that this choice is valid because, in the case of formation of a VC, the maximum space charge of the beam corresponds to the VC region and the center-of-mass oscillations sufficiently accurately describe the features of the VC dynamics.

Below, we offer a more detailed analysis of the regimes in the system in terms of the results presented in Figs. 1 and 2.

Region $T$ (Fig. 1) corresponds to the complete transmission of the electron flux through the drift gap in the case when a VC is not formed and the oscillations in the beam are absent. Line *1* in the map of the regimes corresponds to the critical values of the control parameters (Pierce parameter $\alpha$ and potential difference $\Delta\varphi_{cr}$). Above these values, the system exhibits instability and a nonstationary VC that oscillates in time and space emerges in the beam. In the absence of deceleration ($\Delta\varphi = 0$), the critical value of the Pierce parameter corresponding to formation of a nonstationary VC is $\alpha_{cr} = 4/3$ [13]. This critical value is indicated with arrow on the abscissa axis in Fig. 1. When potential difference $\Delta\varphi$ that leads to deceleration of electrons increases, the boundary of the region in which a nonstationary VC emerges shifts toward smaller values of Pierce parameter $\alpha$. Thus, an increase in the decelerating potential at the second grid of the diode leads to generation of microwave oscillations with the help of a VC at the beam currents that are lower than the currents flowing in a system without deceleration.

---
[2] Hence, in the numerical simulation, we consider the so-called reditron model of a VC-based oscillator, in which the electrons reflected from the VC are eliminated from the interaction process (are deposited at a specially designed anode) [2, 8].

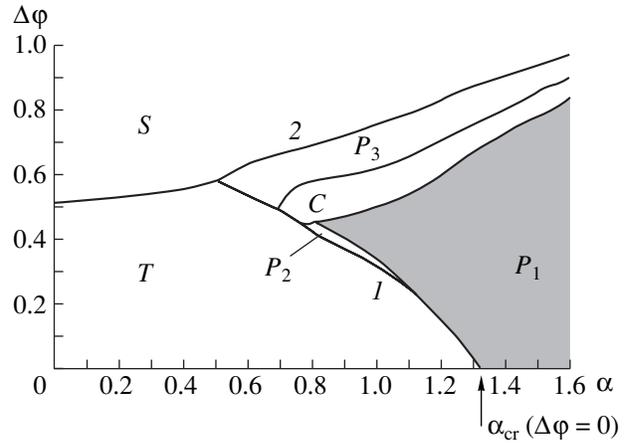

**Fig. 1.** Characteristic regimes of the electron flux in the diode gap with deceleration plotted on the map of parameters (Pierce parameter and decelerating potential difference): (*C*) chaotic oscillations in a beam with a VC, ($P_1$–$P_3$) quasi-periodic and periodic VC oscillations of various types, (*S*) a stationary VC, and (*T*) total transmission of the electron flux through the diode gap. Arrow shows the value of the control parameter $\alpha_{cr} = 4/3$ at which a nonstationary VC is formed in the system in the absence of deceleration ($\Delta\varphi = 0$).

Line *2* in Fig. 1 corresponds to the values of control parameters at which oscillations in the system are suppressed and a stationary VC (from which all the electrons injected into the diode gap are reflected) emerges in the beam. This regime (Fig. 1, region *S*) is observed at a relatively strong deceleration and can be described analytically in the framework of the stationary theory of an electron flux with the supercritical current (see, for example, [30]).

In the region between lines *1* and *2*, as the decelerating potential increases, the system sequentially exhibits oscillation regimes $P_n$ and $C$ in the beam with a VC. Regimes $P_n$ are regular VC oscillations that differ from each other by the phase portrait and the spectral structure. In particular, for low decelerating potentials at the second grid, the system exhibits oscillations $P_1$ and $P_2$. Figures 2a and 2b show the corresponding characteristics of the center-of-mass oscillations for these regimes with low deceleration of the electron beam. It is seen that, in this case, the power spectra represent discrete sets of frequencies against a relatively low noise pedestal. Note that the spectra corresponding to regimes $P_1$ and $P_2$ substantially differ from each other. In regime $P_1$ (Fig. 2b), the fundamental frequency is shifted toward lower frequencies than in regime $P_2$ (Fig. 2a). Note two low-frequency spectral components against a noise pedestal in regime $P_1$. The phase portraits of these regimes differ by the shape and height of the noise pedestal.

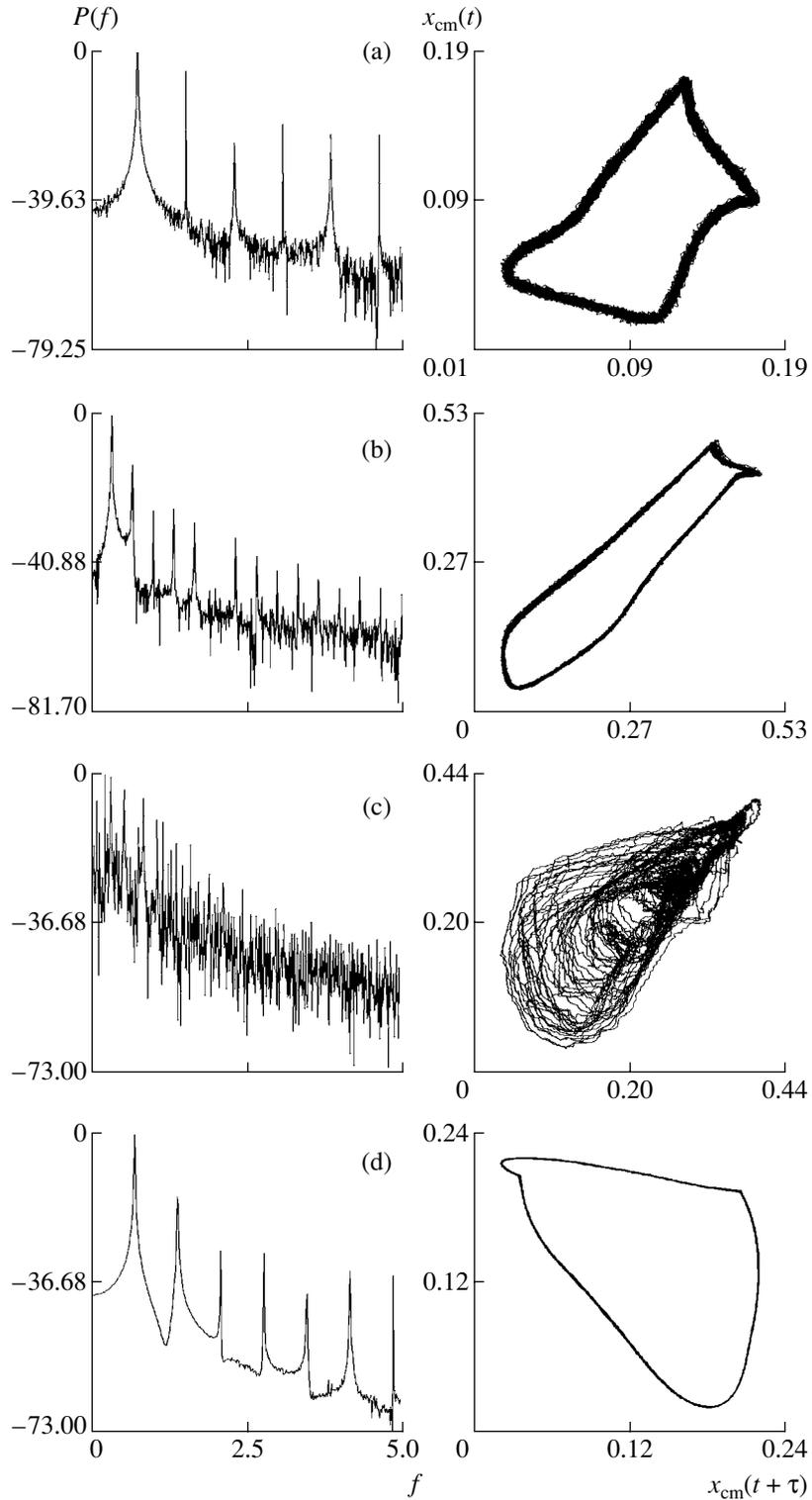

**Fig. 2.** (Left-hand panels) Spectra and (right-hand panels) phase portraits for the control parameter $\alpha = 0.9$ and the decelerating potential $\Delta\varphi =$ (a) 0.37, (b) 0.43, (c) 0.50, and (d) 0.60 (regimes $P_2$, $P_1$, $C$, and $P_3$, respectively).

Then, as the decelerating potential of the second grid increases, the oscillations become more complicated and the system exhibits chaotic oscillations (Fig. 1, regime $C$). Figure 2c shows the characteristics of chaotic VC oscillations. Note the continuous and irregular character of the power spectrum of these oscillations. The phase portrait corresponds to the regime of a developed chaos and represents a chaotic attractor.

At a relatively strong deceleration ($\Delta\varphi > 0.6$), the system once again exhibits regular VC oscillations (Fig. 1, region $P_3$ in the vicinity of line 2). Figure 2d shows the characteristics of regular oscillations corresponding to a relatively strong decelerating potential difference between the grids of the diode gap ($\Delta\varphi = 0.60$). It is seen that, in comparison with regimes $P_1$ and $P_2$, regime $P_3$ corresponds to the absence of a noise pedestal in the spectrum and to a strongly changed shape of the phase portrait.

In the study of the beam with a nonstationary VC, it is important to analyze the power of the observed VC oscillations. In this regard, we consider variations in the power of the VC oscillations as a function of the decelerating potential difference. In the model under study, the power is calculated as a difference of the kinetic energies of electrons that enter the interaction space and the electrons that escape from the diode gap.

Figure 3 demonstrates the dependence of the oscillation power in the beam with a VC on decelerating potential difference $\Delta\varphi$. The halftone zone with the regions corresponding to different oscillation regimes (Fig. 1) corresponds to the VC generation zone. It is seen that, at a low deceleration, the power linearly increases in the regime of regular oscillations. Then, as parameter $\Delta\varphi$ increases, the power reaches a maximum in the chaotic regime and monotonically decreases until the moment when the VC steady state is established in the system.

Thus, it is seen that an increase in the decelerating potential at the control parameter $\alpha > 0.75$ first gives rise to regular VC oscillations. The VC oscillation power increases linearly. Then, as the decelerating potential difference increases, the oscillations become more complicated, their spectral structure and phase portrait vary (Fig. 2b), and the chaotic oscillations emerge (Fig. 2c). With a further increase in the decelerating potential, the VC oscillations are regularized: first, they become periodic (Fig. 2d); then, the oscillations are suppressed and a stationary VC is established in the beam. In the chaotic regime, the power generated in the system reaches the maximum level (Fig. 3) and, then, decreases until the moment corresponding to formation of a stationary VC. Obviously, the cause of the steady state in a beam with a VC is a strong decelerating field that presses the VC toward the injection plane. In this case, the electron flux is reflected completely in the region of the stationary VC and leaves the interaction space through the first grid.

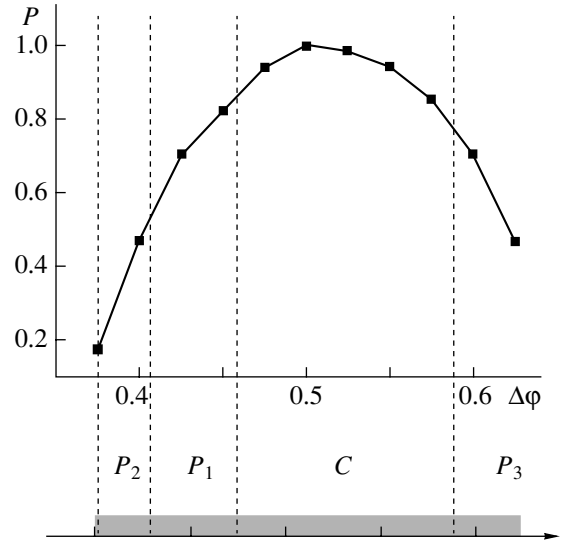

**Fig. 3.** VC oscillation power vs. decelerating potential $\Delta\varphi$ for $\alpha = 0.9$: ($P_1$–$P_3$) regular oscillations corresponding to the values of parameter $\Delta\varphi$ and ($C$) chaotic VC oscillations.

## 3. SELECTION OF COHERENT STRUCTURES AND PHYSICAL PROCESSES IN THE SYSTEM UNDER STUDY

Let us consider physical processes in an electron beam with a VC. For this purpose, we employ a conventional method for the analysis of the complicated dynamics of distributed systems: the method of orthogonal decomposition of spatiotemporal data with the Karhunen–Loéve (KL) expansion [31, 32]. This expansion is helpful during the analysis of the structure formation and complicated dynamics in various distributed systems and in the problems of hydrodynamics, plasma physics, and microwave electronics. The KL method makes it possible to interpret the complicated dynamics of a system with regard to the existence and interaction of space–time structures [9, 12].

The application of this method in the analysis of the complicated behavior of the VC in a beam of charged particles enables one to select space–time structures in the electron flux that have characteristic spatial distributions and time scales in the electron beam (electron bunches) whose interaction can explain features of a beam with a VC [9, 12, 13].

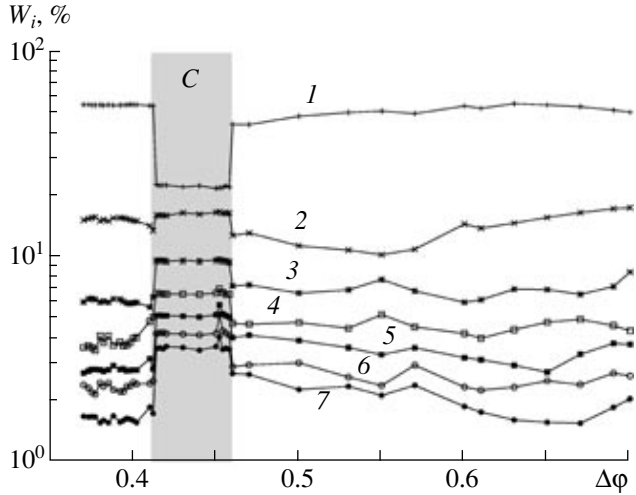

**Fig. 4.** First seven KL modes (curves *1–7*) vs. decelerating potential $\Delta\varphi$ for $\alpha = 0.9$.

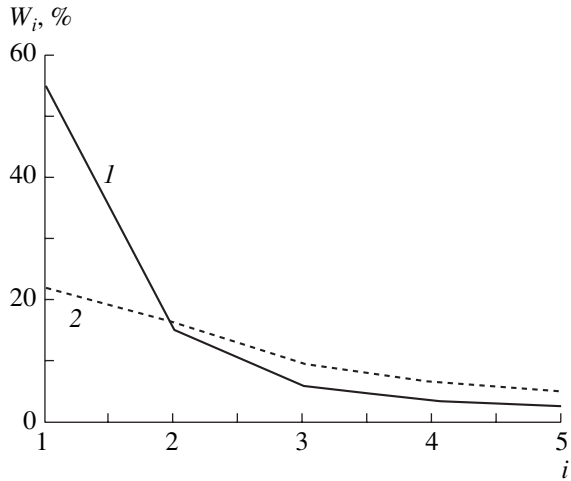

**Fig. 5.** Mode energies $W_i$ vs. mode number *i* for $\alpha = 0.9$ and $\Delta\varphi =$ (*1*) 0.37 and (*2*) 0.43 (oscillations of the first and second type, respectively).

The method of orthogonal decomposition involves solution of the following integral equation:

$$\int K(x, x^*)\Psi(x^*)dx^* = \lambda\Psi(x), \qquad (7)$$

where $K(x, x^*)$ is the equation kernel represented as

$$K(x, x^*) = \langle \xi(x, t)\xi(x^*, t)\rangle_t. \qquad (8)$$

Here, $\langle\ldots\rangle_t$ denotes time averaging. Function $\xi(x, t)$ is the space–time distribution of any physical quantity that is used to analyze the behavior of a system. Prior to the formation of the matrix of kernel $K(x, x^*)$, we need to reduce quantity $\xi(x, t)$ to the zero mean value.

The solution of the problem given by expressions (7) and (8) lies in the search for a set of eigenvalues $\{\lambda_n\}$ and eigenvectors $\{\Psi_n\}$. Each eigenvalue $\lambda_n$ corresponds to a certain eigenvector $\Psi_n$ that determines the *n*th KL mode of the oscillatory process. Quantity $\lambda_n$ is proportional to the energy of the corresponding mode that can be normalized for reasons of convenience as

$$W_n = \frac{\lambda_n}{\sum_i \lambda_i} \times 100\%. \qquad (9)$$

In this study, function $\xi(x, t)$ represents the space–time distribution of space-charge density $\rho(x, t)$ of the electron flux in the diode gap.

Using the orthogonal decomposition of distribution $\rho(x, t)$, we consider oscillation-energy distribution $W_i(\Delta\varphi)$ over the KL modes. Figure 4 demonstrates the dependence of energy $W_i$ of the first seven KL modes on decelerating potential $\Delta\varphi$ at the control parameter $\alpha = 0.9$. These dependences have three characteristic regions: two regions where the mode energy is varied relatively smoothly and one region (halftone region *C*) with stepwise variations in the energy level. Below, we refer to the oscillations that correspond to these regions as the oscillations of the first and second type, respectively.

The fact that the VC oscillation energy is concentrated in a few of the first modes is a characteristic feature of the processes occurring in a beam with a VC. Note that up to 60% of the energy is concentrated in the highest mode (the left-hand region in the figure). The energies of the remaining modes decrease sharply. In particular, the energies of the second and third modes are $W_2 \approx 16\%$ and $W_3 \approx 6\%$, respectively, etc.

For the decelerating potential range $\Delta\varphi \simeq 0.41$–0.46, we observe a modification of the inner structure of the electron flux that lies in a stepwise energy redistribution between the highest and lowest KL modes. As the flux deceleration is varied, the energy of the first mode jumps from 60 to 20%, while the energy of the lower modes increases. Figure 5 demonstrates the dependence of mode energy $W_i$ on mode number *i* for oscillations of the first (curve *1*) and second (curve *2*) types. It is seen that the energy of the highest mode is redistributed to the modes of lower orders. Note that the number of structures with a relatively large energy ($W_i > 5\%$), which are important for the oscillation process, increases also. The region of the oscillations of the second type that corresponds to the energy redistribution between the highest and lowest harmonics completely coincides with the halftone beak in Fig. 1.

The third characteristic region (on the right-hand side in Fig. 4) corresponds to the parameters of the chaotic ($\Delta\varphi = 0.46$–0.58) and periodic ($\Delta\varphi = 0.580$–0.725) oscillations. In this case, the energy distribution over modes does not exhibit substantial variations in various oscillation regimes.

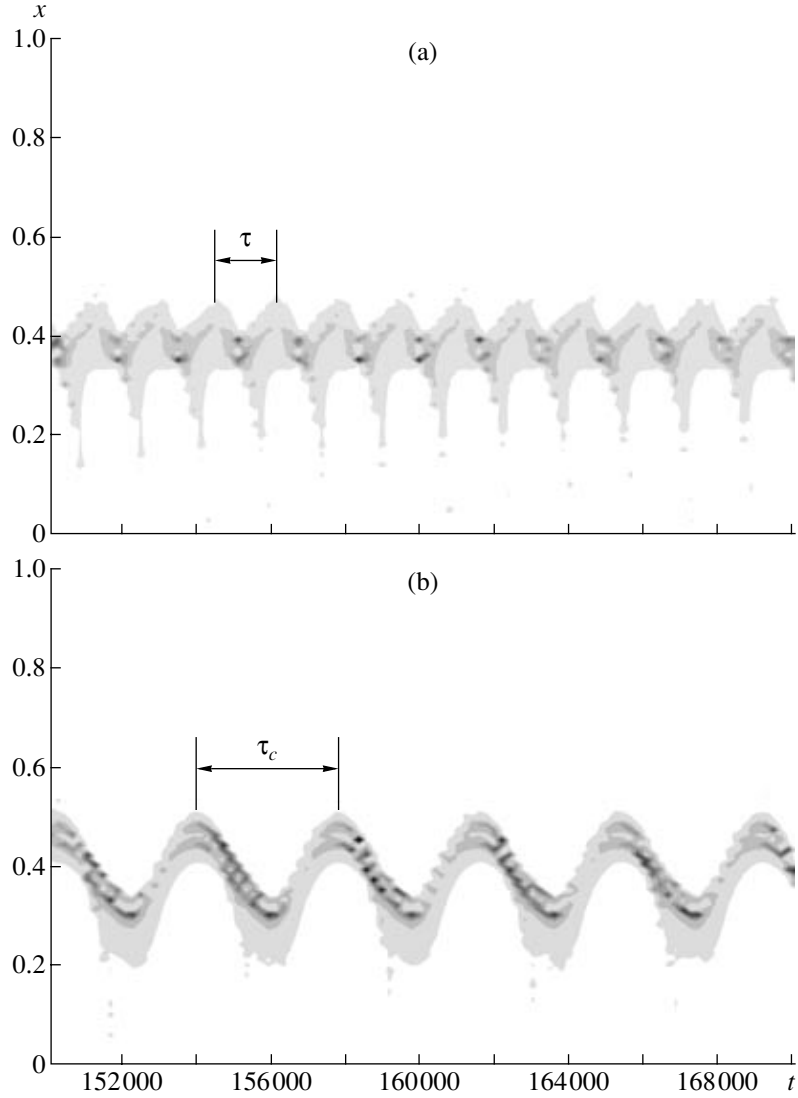

**Fig. 6.** Time distribution of quantity $\rho(x, t)$ in the interaction space at $\alpha = 0.9$ and $\Delta\varphi =$ (a) 0.37 and (b) 0.43 (oscillations of the first and second type, respectively). Shades of gray show the charge density: white corresponds to the lowest density and dark gray and black correspond to the highest density (a VC formed in the interaction space).

To interpret the physical processes that accompany changes in the VC oscillation regimes with variations in the decelerating potential, we consider variations in the behavior of space-charge density $\rho(x, t)$ in space and time for the oscillations of the first and second types. Figure 6 demonstrates the corresponding distributions. Shades of gray indicate the space-charge density at each point at a given moment. Figures 6a and 6b show the distributions for the decelerating potentials $\Delta\varphi = 0.37$ and 0.43, respectively, that correspond to the oscillations of the first and second types. It is seen that the spatial distributions have different space–time structures in two different regimes characterized by significantly different energy distributions over modes. In the first case, the electron bunches are localized in a certain spatial region in the vicinity of the point $x = 0.4$. In the course of time, the charge is periodically delivered to the first accelerating grid with characteristic time $\tau$ (Fig. 6a). In the case shown in Fig. 6b, the VC performs periodic spatial oscillations around the point $x \approx 0.4$ with a relatively large amplitude rather than remaining localized in a relatively small area of the interaction space. Characteristic time $\tau_c$ of the VC oscillations is approximately twice as long as $\tau$ (Fig. 6a).

The dynamics of the electron bunches in the VC area can be easily traced with the aid of the corresponding space–time diagrams of the electron flux (Fig. 7). Figures 7a and 7b show these diagrams with the trajectories of the charged particles for the regimes shown in Fig. 6. As the beam moves from the first grid ($x = 0$) to the second grid ($x = 1$) in the presence of a decelerating potential, the charge is accumulated between the grids in the diode gap. Thus, the potential of the intergrid space decreases. A VC is formed at the moment when the potential of a certain area becomes equal to zero. When charges are accumulated in the VC area, the potential barrier becomes higher, a minor drift of the electron bunch takes place, and the charge is dropped toward the first grid with a positive potential relative to the VC. Thus, the potential barrier becomes lower and the electrons can cross the gap between the grids. This process is periodically repeated with a certain repetition rate. In the region corresponding to oscillations of the second type, the process is qualitatively different. In this case, the VC periodically oscillates in space around some mean value. The repetition rate of the process decreases by a factor of approximately 2.

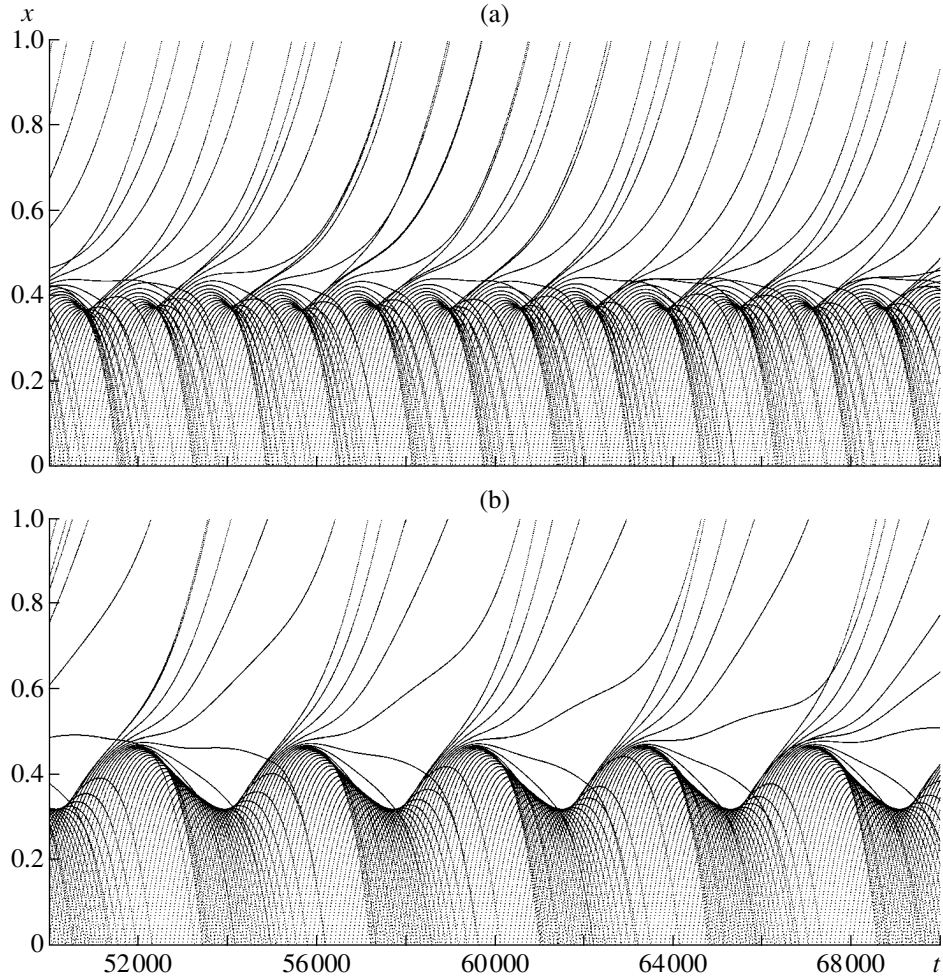

**Fig. 7.** Trajectories of charged particles in the diode gap plotted for $\alpha = 0.9$ and $\Delta\varphi = $ (a) 0.37 and (b) 0.43 (oscillations of the first and second type, respectively).

To study physical processes in a beam with a VC, we need such characteristic of the beam as the velocity distribution of charged particles. Figure 8 shows velocity distributions for the regimes of the first and second types. This distribution was determined in the spatial plane with the coordinate $x = 0.3$. The coordinate at which we measure the electron velocity distribution is such that the corresponding spatial plane is located between the first grid and the region of VC oscillations and does not belong to the latter. It is seen that different velocity distributions of the electron flux correspond to oscillations of the first and second type. In the case of oscillations of the second type, the velocity distribution is substantially wider. A possible reason for this circumstance is that a relatively large amplitude of the spatial VC oscillations in the second regime leads to a significant enhancement of the effect of the potential barrier on the velocity of the electron beam in the vicinity of the first grid. This phenomenon results in a substantial velocity modulation of the electron flux in the vicinity of the first grid. Let us discuss this effect in more detail.

A forced modulation of the electron flux in the oscillation regime of the first type makes it possible to demonstrate the strength of the effect of VC oscillations on the electron-flux velocity in the vicinity of the first grid for the oscillation regime of the second type. Let the electron-flux modulation in the vicinity of the first grid be represented as

$$v = v_0(1 + m\sin(2\pi f_c t)), \qquad (10)$$

where $v_0$ is the unperturbed velocity of the electron flux, $m$ is the modulation depth, and $f_c = 1/\tau_c$ is the frequency that corresponds to the VC oscillation frequency in the oscillation regime of the second type (Fig. 7b).

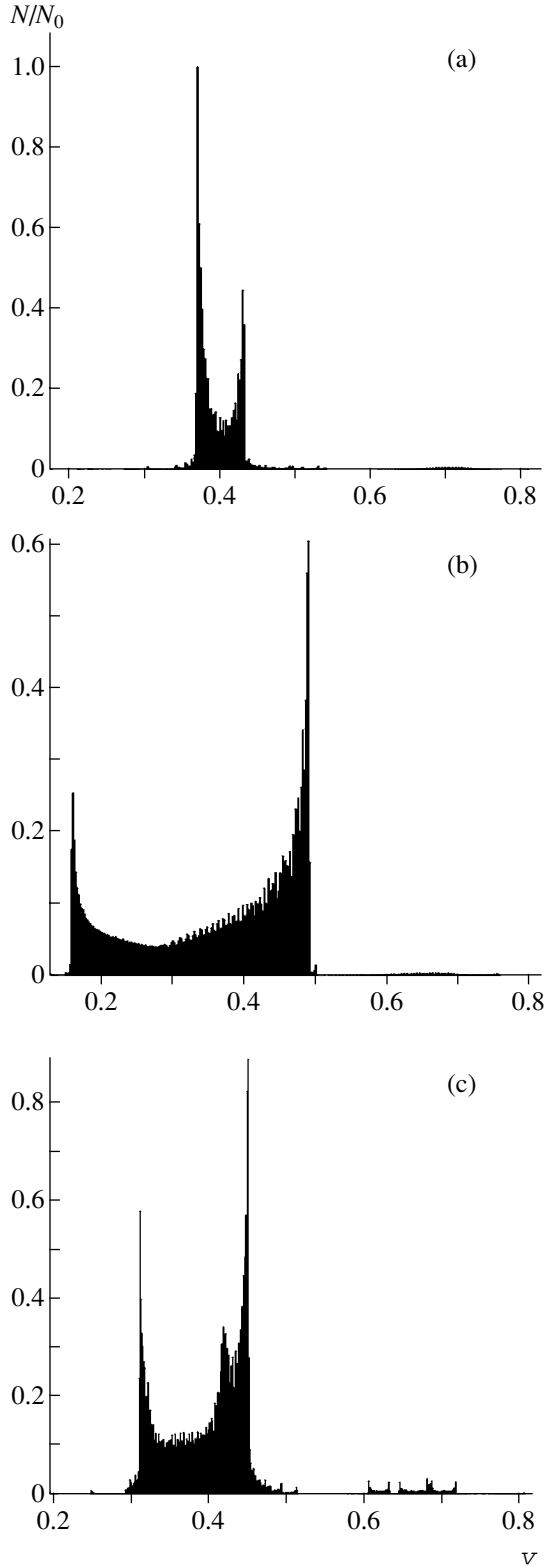

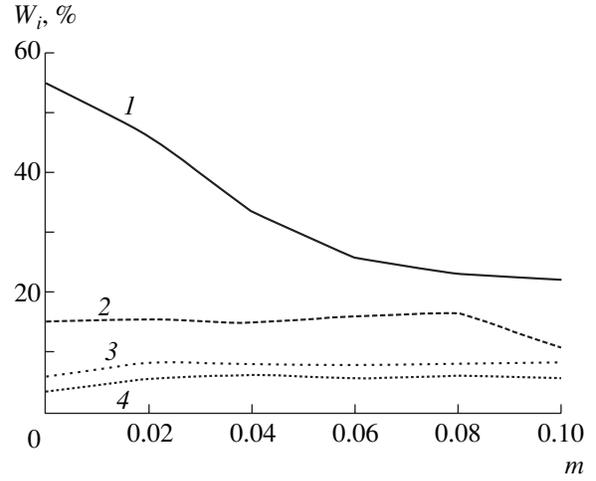

**Fig. 9.** Energies of modes *1–4* vs. the modulation depth at $\Delta\varphi = 0.38$.

Figure 8c shows the velocity distribution for the oscillations of the first type in the presence of additional forced modulation (10). It is seen that, in the case of a periodic modulation, the velocity distribution broadens and becomes approximately twice as broad as the velocity distribution plotted for the same decelerating potential in the absence of modulation (Fig. 8a).

To corroborate the data obtained, we calculated the dependence of the KL-mode energy on the modulation depth of the electron-flux velocity (Fig. 9). It is seen that an increase in the modulation depth causes a decrease in the energy of the first mode and an increase in the energies of lower modes. The same behavior of the mode energies is observed in the case of oscillations of the second type (Fig. 4).

Thus, on the basis of the electron velocity distributions and the dependence of the mode energy on the modulation depth, we draw the conclusion that the electron-flux velocity is strongly modulated in the vicinity of the first grid in the case of oscillations of the second type. This modulation is not observed for oscillations of the first type. The VC oscillations become more complicated, and the number of significant KL modes increases with an increase in the modulation depth.

**Fig. 8.** Velocity distributions in the direction of the electron motion at the point $x = 0.3$: (a) in the absence of modulation at $\Delta\varphi = 0.38$ and $N_0 = 33\ 839$; (b) in the absence of modulation at $\Delta\varphi = 0.43$ and $N_0 = 18\ 281$; and (c) in the presence of the electron-flux modulation for the modulation depth $m = 0.025$ at the input, $\Delta\varphi = 0.38$, and $N_0 = 19\ 793$.

## 4. EXPERIMENTAL STUDY OF THE PROCESSES IN A BEAM WITH A VIRTUAL CATHODE IN THE PRESENCE OF A DECELERATING FIELD

To corroborate the results of numerical simulation of the nonlinear dynamics of an electron beam with a VC in the presence of a decelerating field, we experimentally studied oscillations in a beam with a VC. In the experiments, we used a prototype of the diode scheme similar to the scheme used in the numerical simulation. In the experimental setup, the electron beam generated by an electron-optical system (EOS) is injected into a diode that consists of two grid electrodes with a decelerating field (Fig. 10a). The decelerating field is generated owing to negative potential $V$ of the output (second) grid relative to the input (first) grid.

A hot cathode that works in the space-charge-limited mode serves as a source of electrons. The EOS forms an axially symmetric converging cylindrical electron beam. In the experiments, the accelerating voltage of the electron beam is 1.5 kV. The beam current at the EOS exit (100 mA) is determined by the cathode heating. The electron-velocity spread at the EOS exit is relatively low (no greater than 0.2% in relative units), so that the electron beam is assumed a monovelocity beam.

The electron beam from the EOS exit arrives at the grid (diode) gap. Potential $V_0$ of the first grid is equal to anode potential $V_a$ (accelerating voltage), whereas the potential of the second grid $V_d = V_0 - \Delta V_d$ ranges from $V_d/V_0 = 1$ ($\Delta V_d = 0$, in the absence of deceleration) to $V_d/V_0 = 0$ ($\Delta V_d = V_0$, in the case of the complete deceleration of the electron flux). Quantity $\Delta V_d$ is the potential difference between the grids that generates the static decelerating field in the diode gap.

When decelerating potential difference $\Delta V_d$ between the grids of the diode gap increases, we observe formation of a VC in the electron flux. The VC oscillations in time and space modulate the electron beam, so that electrons are partially reflected from the VC toward the input grid. Thus, chaotic VC oscillations emerge in the system. The shape and power of these oscillations substantially depend on potential difference $\Delta V_d$ between the grids of the diode gap.

To analyze the noiselike oscillations in the electron beam, we employ a broadband fragment of a helical slow-wave structure (FHSWS) loaded with an absorbing insert and an energy output as well as a fragment of a wideband coaxial line loaded with a capacitance (HF probe) that can be translated along three perpendicular axes [22]. The coaxial probe is supplemented with a decelerating grid for the spectral analysis of the longitudinal velocities of charged particles. In the diode gap, the velocity- and density-modulated electron beam excites the FHSWS, whose signal is studied with an SCh-60 spectrum analyzer. This technique makes it possible to determine the spectral power density of the noise of oscillations generated by an electron beam with a VC.

The experiments were performed on an assembled vacuum setup (Fig. 10b) under continuous evacuation with a minimum residual gas pressure of $10^{-7}$–$10^{-6}$ Torr.

As was mentioned, the oscillations in the system under study are determined by the presence of a VC in the intergrid space with a decelerating field. The exper-iments have shown that, at a relatively low value of decelerating potential difference $\Delta V_d/V_0$, the oscillations are not observed in the electron beam. When deceleration of the beam electrons increases ($\Delta V_d/V_0$ increases), a VC is formed in the system at a certain critical level of the potential difference, $[\Delta V_d/V_0]_{cr}$. Electrons are partially reflected from the VC toward the first grid of the diode. At low deceleration, the VC oscillations are nearly regular and the spectrum of the generated radiation is discrete. Wideband noiselike oscillations are generated in the system when deceleration of electrons is increased. At a relatively high deceleration, generation vanishes again.

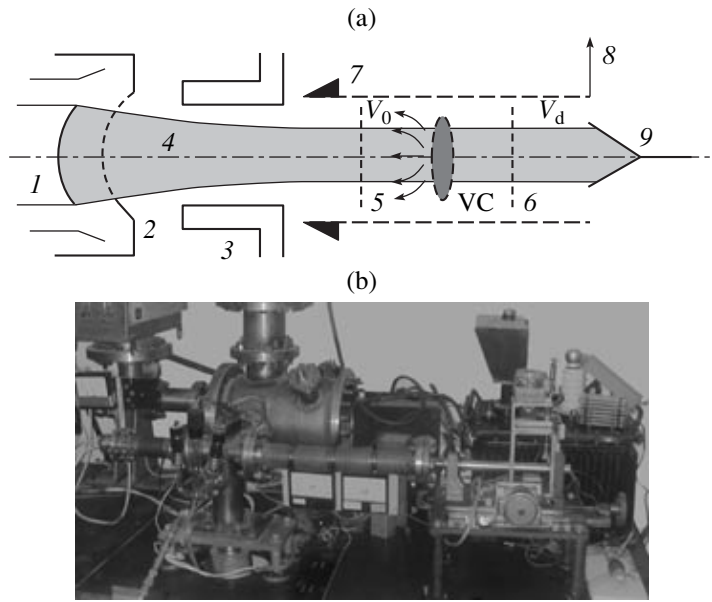

**Fig. 10.** (a) Design and (b) photograph of the experimental setup for studying chaotic oscillations in a beam with a VC: (*1*) the hot cathode, (*2*) the grid of the gun, (*3*) the second anode of the gun, (*4*) the converging electron beam, (*5*) the input grid of the diode gap, (*6*) the output grid of the diode gap with a decelerating potential, (*7*) FHSWS, (*8*) the energy output, and (*9*) the collector.

Figure 11 demonstrates the power spectra of the microwave oscillations that were measured experimentally at various decelerating potentials. It is seen that an increase in deceleration leads to a gradual complication of the power spectrum. As the decelerating potential increases, the system exhibits various regimes. First, we observe single-frequency oscillations that are characterized by a relatively narrow spectrum (Fig. 11a). Then, as this parameter increases, the oscillations become multifrequency and the spectrum becomes more complicated but remains discrete at low deceleration levels (Figs. 11b, 11c). The VC oscillation frequency increases. When deceleration further increases (while $\Delta V_d/V_0$ increases), the oscillations become chaotic at certain potentials of the second grid. The spectrum becomes continuous in a relatively wide frequency range. This scenario of the development of chaotic oscillations is in reasonable agreement with the theoretical analysis of the processes in the diode gap in the presence of a decelerating field.

Figure 12a shows the dependence of the normalized integral oscillation power on potential difference $\Delta V_d/V_0$ between the grids at the electron velocity spread $\Delta v/v_0 \sim 0.2\%$. It is seen that, at low deceleration, integral power $P_\Sigma$ of the oscillations in the beam with a VC is relatively small. When $\Delta V_d/V_0$ increases, the power increases and reaches its maximum value at a certain optimal decelerating potential. Then, the integral oscillation power decreases. The experimental dependence is in qualitative agreement with the theoretical dependence (Fig. 3). In addition, Fig. 12a shows quantity $K$ (the electron-beam transmission efficiency), determined as the ratio of output current $I_{out}$ of the second grid to beam current $I_0$ of the electron gun: $K = I_{out}/I_0$. It is seen that an increase in the decelerating potential leads to a monotonic decrease in $K$ (in contrast to integral oscillation power $P_\Sigma$). This circumstance means that an increase in decelerating potential difference $\Delta V_d$ causes an increase in the number of electrons reflected from the VC or in the diode gap behind the VC toward the input grid. At a certain (relatively high) potential difference $\Delta V_d$ between the grids, we observe almost total reflection of electrons from the VC, so that the electron-beam transmission becomes close to zero ($K \approx 0$). In this regime, the VC in the electron beam becomes stationary and generation vanishes (the integral power equals zero).

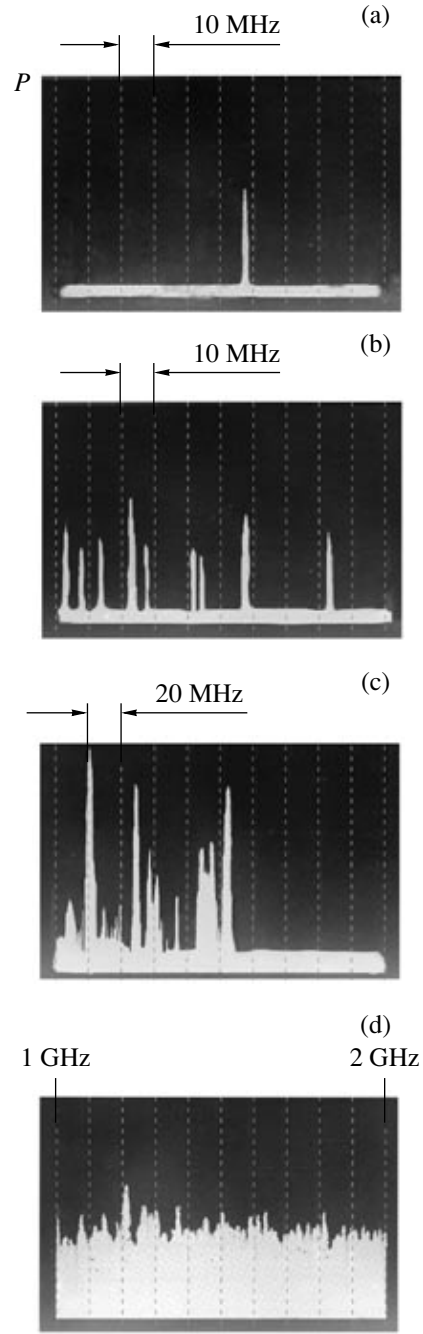

**Fig. 11.** Experimental power spectra of the VC oscillations obtained at $\Delta V_d/V_0 =$ (a) 0.20, (b) 0.24, (c) 0.28, and (d) 0.50.

Figure 12b demonstrates the regions of the characteristic regimes of oscillations for a VC system with a monovelocity beam under a variation (increase) in the decelerating potential. The halftone region corresponds to formation of a nonstationary VC (generation of microwave radiation). The diagram shows the range of the decelerating potential in which the oscillations in the beam with a VC are absent (region $S$), the region of regular VC oscillations (region $P$), and the region of chaotic wideband generation (region $C$). It is seen that the system demonstrates a transition from the steady state to chaotic oscillations through the region of aperiodic dynamics. Note that an increase in the decelerating potential results in complication of chaotic oscillations (broadening of the generated frequency band and regularization of the power spectrum).

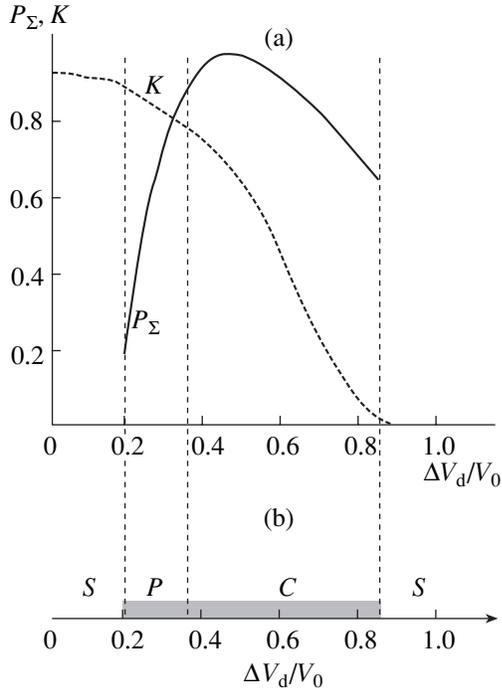

**Fig. 12.** (a) Normalized integral power $P_\Sigma$ of interaction and electron-beam transmission efficiency $K$ of the diode vs. normalized potential difference $\Delta V_d/V_0$ between the grids for a monovelocity beam and (b) regions of characteristic generation regimes of a VC system with a monovelocity beam vs. the decelerating potential: ($S$) the absence of a microwave generation, ($P$) regular VC oscillations, and ($C$) chaotic wideband generation. The halftone region corresponds to formation of a nonstationary VC (microwave generation in the electron–wave oscillator under study).

Thus, the experimental data qualitatively confirm the results obtained during numerical simulation of the nonlinear dynamics of an electron flux with a VC. The processes occurring in the beam after formation and interaction of electron structures observed in the numerical simulation need to be studied additionally in experiments.

## CONCLUSIONS

The nonlinear nonstationary processes in an electron beam with a VC in the diode gap with a decelerating potential have been studied theoretically and experimentally. It has been demonstrated that the system exhibits versatile dynamics (from single-frequency oscillations to developed wideband chaotic VC oscillations) under variations in the control parameters (primarily, the decelerating potential). The regions that correspond to various regimes of VC oscillations have been plotted on the map of parameters (decelerating potential $\Delta \varphi$ and Pierce parameter $\alpha$). Using the Karhunen–Loéve expansion (orthogonal decomposition), we have selected various types of the VC behavior in the presence of a decelerating field and have demonstrated that these types are characterized by different types of space–time dynamics in the electron beam with a VC. The analysis of the physical processes has shown that variations in the VC dynamics are determined by an increase in the electron-velocity modulation at the VC oscillation frequency in the vicinity of the input grid.

A distinctive feature of the system under study is the possibility of effectively controlling oscillation regimes in the system with the help of a variation in decelerating potential difference $\Delta \varphi$ in the diode gap. Note the possibility for tuning of the system from the regime of regular oscillations with a narrow spectrum to the regime of chaotic oscillations with a wideband generation and a uniform spectrum.

The experimental study of a prototype of this physical system shows that the proposed numerical model yields a sufficiently accurate qualitative (and, partly, quantitative) interpretation of the features of a real system with a VC in the diode gap with a decelerating potential.

## ACKNOWLEDGMENTS

This study was supported by the Russian Foundation for Basic Research (project nos. 05-02-16286 and 05-02-08030), the Program for the Support of the Leading Scientific Schools (grant no. NSh-4167.2006.2), the US Civilian Research and Development Foundation (grant REC-006), and the Dynasty Foundation for Noncommercial Programs.